\documentclass[preprint,aps,amsmath]{revtex4}
\usepackage{graphicx}
\begin{document} 
\preprint{}

\title{Correlations at a quantum phase transition in interacting Bose systems}
\author{Min-Chul~Cha}
\affiliation{Department of Applied Physics, Hanyang University, Ansan,
Kyunggi-do 426-791, Korea}
\date{today}

\begin{abstract}
We have investigated the correlation functions of interacting bosons
at the generic superfluid-insulator transition,
a prototypical quantum phase transition,
in two dimensions in the spherical limit.
Unexpectedly the spatial correlation functions show non-power-law behavior
consisting of two parts: short-range
correlation due to the particle-hole pair excitations
and long-range off-diagonal order due to the single-particle condensation.
The temporal correlation functions, on the other hand, show power-law behavior.
\end{abstract}

\maketitle

Quantum phase transitions\cite{Sachdev}
of interacting bosons\cite{Fisher89}
have continuously drawn significant attention.
They have been realized in many different systems,
such as Josephson-junction arrays\cite{Fazio01},
thin film superconductors\cite{Goldman98},
$^4$He films\cite{Crowell93}
and ultracold atoms in optical lattices\cite{Greiner,Jaksch98}.
Typically the transition occurs between a superfluid and an insulator,
and often serves as a prototype of the quantum phase transition\cite{Sondhi}
because of its convenience in describing the transition in terms of
the establishment of a macroscopic phase coherence.

Quantum phase transitions are frequently discussed in
equivalent classical models which include fluctuations in
the temporal direction.
For the special case with the dynamical critical exponent $z=1$,
the equivalence of the spatial and temporal directions
directly confirms the validity of this mapping.
In this scheme the properties of quantum phase transitions in
$d$ dimensions are readily derived from the classical models
in $d+1$ dimensions.
In general, however, asymmetry between the spatial and temporal directions
in quantum phase transitions results in $z \ne 1$.
Even for this case, a simple extension of the above scheme into $d+z$ dimensions
is frequently assumed.
Many scaling ansatz, such as hyperscaling relations\cite{Kim91},
are constructed on the basis of this assumption.
Similarly, correlation functions at the transition
are assumed to have power-law behavior effectively in $d+z$ dimensions.
Some numerical works have used this assumed property to extract
the value of $z$ from correlation functions\cite{Wallin94,corr}.
However, it is quite plausible that the asymmetry of fluctuations
in the spatial and the temporal directions, which brings non-unity
of $z$, would modify the correlation functions.

In this work, we investigate the scaling properties of the quantum
rotor model, which is equivalent to the boson Hubbard model in
large density limit, at the generic superfluid-insulator transition
in two dimensions in the spherical limit.
We find that the scaling behavior clearly supports hyperscaling
in $d+z$ dimensions with $z=2$, as predicted\cite{Fisher89}.
However, the spatial correlation functions at the transition show,
instead of power-law behavior,
properties consisting of two parts: one is a short-range
contribution from the particle-hole pair excitations
and the other is the off-diagonal long-range order due to single-particle condensation.
This implies that the macroscopic phase coherence is achieved through
the single-particle condensation rather than the establishment of
long-range correlation of fluctuations in a spatial direction,
and we have the background particle-hole excitations as a normal fluid,
instead of the single-particle excitations,
even in the case with a charge-offset.
The temporal correlation functions, on the other hand, show the power-law behavior 
as expected.

The universal features of quantum phase transitions in interacting Bose systems
are divided into two universality classes\cite{Fisher89}
depending on whether the transition is driven either by the phase fluctuations 
or by the density fluctuations of bosons.
At a commensurate density
the particle-hole symmetry is sustained across the transition,
and the transition is driven by the phase fluctuations or,
equivalently, by the particle-hole pair excitations.
This transition belongs to the ($d$+1)-dimensional classical $XY$
spin model with $z=1$.
In the presence of a charge-offset, however,
either single particle or hole excitations are favored:
the transition is marked by the disappearance of
the energy gap for single particle or hole excitations.
Therefore, the universality class of this transition,
known as the generic superfluid-insulator transition,
is characterized by the single-particle nature,
carrying $z=2$ and the mean-field correlation length critical exponent $\nu=1/2$.
It is interesting to investigate whether correlation functions 
could reveal directly the different nature of fluctuations 
in two universality classes.

The essential properties of strongly correlated
interacting bosons can be captured by a boson Hubbard model
\begin{eqnarray}
H={U \over 2} \sum_i  n_i ^2 -\mu \sum_i n_i
- t \sum_{<ij>} (b_i^\dagger b_j + b_j^\dagger b_i ),
\end{eqnarray}
where $b_j (b_j^\dagger)$ is the boson annihilation(creation) operator
at $j$-th site, and $n_j$ is the number operator.
$U$ and $t$ stand for the strength of on-site repulsion and
nearest neighbor hopping respectively,
and $\mu$ is the chemical potential.
It is convenient to put $\mu/U \to n_0 + {\mu}/U$ 
with an integer $n_0$ to make $-1/2 < \mu/U < 1/2$.
Here $n_0$ is an integer representing the background number of
bosons per site.
We study the phase transition of this model on two-dimensional square lattices.

The phase transition of this model is characterized by establishment of
the phase coherence of the order parameter.
In the limit $n_0 \gg 1$, the Hamiltonian is reduced to the quantum rotor model
\begin{eqnarray}
H={U \over 2} \sum_i  n_i ^2 -\mu \sum_i n_i
- 2J \sum_{<ij>} \cos(\theta_i - \theta_j),
\end{eqnarray}
where $\theta_j$ is the phase angle of the order parameter,
$n_j = {-i}{\partial / \partial \theta_j}$, and $J=n_0 t$.

Now we use the spherical approximation to investigate
the quantum phase transition of the above model.
The spherical approximation\cite{Ma} has been widely used 
in studies on phase transitions,
which is very powerful in handling models beyond the mean-field level.
This method treats phase transitions in a scheme which is exact
in the limit that the number of order parameter components goes to infinity.
It has been used to study
quantum phase transitions in Bose systems\cite{Tu94}.
Previously this method was used\cite{Cha00} to study the
superfluid-insulator quantum transition at a commensurate density
in two dimensions, yielding $\nu=1$ and $z=1$, as expected.

Through a path integral mapping, we can construct the corresponding 
classical action
\begin{eqnarray}
S_0[\psi]=-\int^{\bar\beta}_0 d\tau [K \sum_i \psi_i^*(\tau)
(\partial_\tau - \mu)^2 \psi_i(\tau)
 - \sum_{i,j} (\delta_{ij} \sigma-J_{ij}) \psi_i^*(\tau) \psi_j(\tau) ],
\label{eq:S_0}
\end{eqnarray}
where $K = 1/(2U)$, $\bar\beta$ denotes the inverse temperature,
$J_{ij}$ denotes the nearest neighbor hopping matrix elements, 
and $\psi_j=e^{-i\theta_j}$.
The trace in the partition function
should be taken with the constraint $|\psi_j| =1$ for every $j$.
But we introduce a Lagrange multiplier $\sigma$ to
replace the constraint by a less strict self-consistent condition
$1=\langle \psi_j(\tau) \psi_j^*(\tau) \rangle_0$,
where $\langle ... \rangle_0$ represents the average over $S_0$,
which is called the spherical approximation.

In this work we focus on the case with a charge-offset ($\bar n =\mu/U \neq 0$).
Scaling theories predicted that the dynamical critical exponent $z=2$\cite{Fisher89}.
To confirm this through the hyperscaling hypothesis in $d+z$ dimensions,
we investigate the finite-size scaling behavior of the superfluid stiffness
\begin{eqnarray}
\rho_s=L^{-(d+z-2)}X_\rho (L^{1/\nu}(K-K_c), \bar\beta/L^z),
\label{eq:rho0}
\end{eqnarray}
where $X_\rho$ is a scaling function.
By diagonalizing the equation
$\sum_j (\sigma \delta_{ij} -J_{ij}) \phi_j^{\vec q}
= \epsilon^{\vec q} \phi_i^{\vec q}$ to have
$\phi_i^{\vec q}=(1/\sqrt{L^2}) e^{i \vec q \cdot \vec R_i}$
and 
$\epsilon^{\vec q}=\sigma -2J (\cos q_x+\cos q_y)$, where
$\vec R_j$ is the position vector of $j$-th site and
$\vec q = (2\pi/L)(n_x,n_y)$ is the wave vector with integer $n_x$ and $n_y$,
we can easily derive the formula
\begin{eqnarray}
\rho_s={1 \over L^2}
\sum_{\vec q}\Big\{{J \over 2\sqrt{K\epsilon^{\vec q}}}
\big(\cos q_x -{J \sin^2 q_x \over \epsilon^{\vec q}}\big)
\big[\coth{\bar\beta \over 2}(\sqrt{\epsilon^{\vec q}/K}-\mu)
+\coth{\bar\beta \over 2}(\sqrt{\epsilon^{\vec q}/K}+\mu)\big]\nonumber \\
-{J^2 \bar\beta \sin^2 q_x \over 4 \epsilon^{\vec q}K}
\big[{1\over \sinh^2 {\bar\beta \over 2}(\sqrt{\epsilon^{\vec q}/K}-\mu)}
+{1\over \sinh^2 {\bar\beta \over 2}(\sqrt{\epsilon^{\vec q}/K}+\mu)}\big]\Big\}.
\end{eqnarray}
Here we set the lattice constant to be 1, and take the energy unit $J=1$.

\begin{figure}
\center
\includegraphics[width=2.9in]{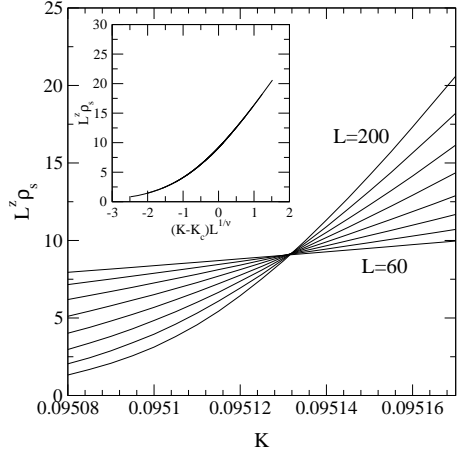}
\caption{The finite-size scaling behavior of the superfluid stiffness,
$\rho_s$ for $\bar n=0.1$.
High quality crossing behavior at $K_c=0.09513(2)$
with $z=2$ for $L=60, 80, 100,..., 200$ is obtained without suffering from
any statistical errors.
We set $\bar\beta/L^z=0.02$. Inset: The data
collapse onto a single curve as a function of the scaling variable $L^{1/\nu}(K-K_c)$  
with $\nu=0.5$.}
\label{fig:rho0}
\end{figure}

Figure~\ref{fig:rho0} shows the finite-size scaling behavior of $\rho_s$ 
for $\bar n = 0.1$.
With $z=2$, curves for different sizes $L=60,80,100,...,200$ cross at a point
$K_c=0.09513(2)$.
Here we fix the aspect ratio to have $\bar\beta/L^z=0.02$, and obtain
the correlation critical exponent $\nu=0.5$, as shown in the inset.
The high quality crossing behavior implies that $\rho_s L^{(d+z-2)}$ is indeed
a universal quantity in the vicinity of the critical point,
strongly supporting the hyperscaling hypothesis in $d+z$ dimensions.

Now we turn our attention to the correlation functions.
The correlation function between $\vec R_i$ and $\vec R_j$
with distance $\tau$ in time is given by
\begin{eqnarray}
G(|\vec R_i-\vec R_j|, \tau)
={1 \over L^2}\sum_{\vec q}{e^{i \vec q \cdot (\vec R_i -\vec R_j)}
\over 2 \sqrt{K\epsilon^{\vec q}}}
\big[{e^{-(\bar\beta-\tau) (\sqrt{\epsilon^{\vec q}/K}+\mu)} \over
  1-e^{-\bar\beta (\sqrt{\epsilon^{\vec q}/K}+\mu)}}
+{e^{-\tau (\sqrt{\epsilon^{\vec q}/K}-\mu)} \over
  1-e^{-\bar\beta (\sqrt{\epsilon^{\vec q}/K}-\mu)}}\big],
\end{eqnarray}
for $0<\tau<\bar\beta$.
It was assumed\cite{Fisher89,Wallin94} that at the critical point
the correlation functions show long-range power-law behavior
in $d+z$ dimensions so that
$G(|\vec R_i - \vec R_j|, \tau) \sim 1/(|{\vec R_i}-{\vec R_j}|^2+\tau^{2/z})^{y/2}$,
where $y=d+z-2+\eta$ ($\eta$ is the anomalous critical exponent).
Based on this assumption, estimation of the value of $z$
has been made\cite{Wallin94,corr,Alet04} by comparing the asymptotic behavior
of the spatial and temporal correlation functions.

\begin{figure}
\center
\includegraphics[width=2.9in]{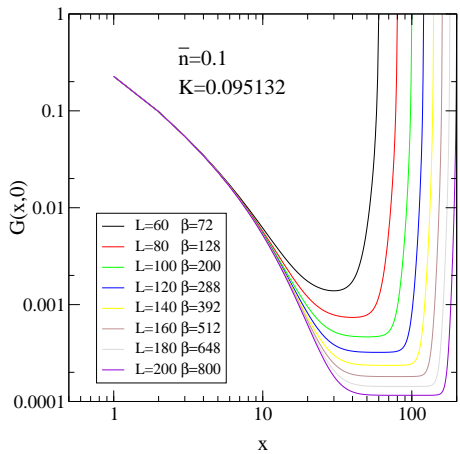}
\caption{Log-log plot of the correlation functions along a spatial axis 
at the critical point $K_c=0.095132$ for $\bar n=0.1$.
They do not show the power-law behavior.
The curves are fitted perfectly onto the form
$G(x,0)=A (x^{-1}e^{-x/\lambda}+(L-x)^{-1}e^{-(L-x)/\lambda})  + C$.
}
\label{fig:gx}
\end{figure}

Figure~\ref{fig:gx} shows the spatial correlation functions along a spatial
direction at the critical point.
Surprisingly they do not show the power-law behavior;
instead, the correlation functions have forms
\begin{eqnarray}
G(x,0)=A (x^{-1}e^{-x/\lambda}+(L-x)^{-1}e^{-(L-x)/\lambda})  + C
\end{eqnarray}
for $1 \ll x \ll L$,
where $\lambda$ is a finite number,
insensitive to the size of the system, $C$ is a constant depending on
the size, and $A$ is a fitting parameter. For example, for the curve
in Fig.~\ref{fig:gx} with $L=200$, $A=0.26$, $\lambda=6.11$, and $C=1.17\times 10^{-4}$.
Similar features found in Monte Carlo calculations were
discussed as a result of finite-size effect\cite{Alet04}.
Here we use the same value of the aspect ratio $\bar\beta/L^z=0.02$,
previously used in Fig.~\ref{fig:rho0}.
For significantly wide range of $\bar\beta/L^z$, the same behavior is found.
Very small aspect ratios, however, change the behavior, possibly due to 
finite-temperature effects, but still no power-law behavior occurs.

\begin{figure}
\center
\includegraphics[width=2.9in]{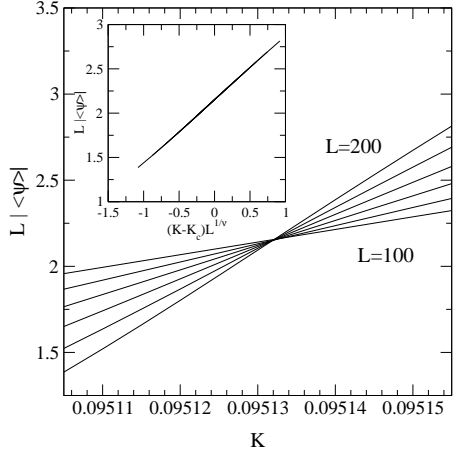}
\caption{The finite-size scaling behavior of the off-diagonal long-range order parameter
for $L=100,120,...,200$, providing the mean-field critical
exponents $\nu=0.5$ and ${\beta}/\nu=1$. 
}
\label{fig:offd}
\end{figure}

We can easily identify the meaning of the constant $C$. The spatial correlation
function is actually flat for $x \approx L/2$. In other words,
$C=G(L/2,0)=|\langle\psi\rangle|^2$, where $|\langle\psi\rangle|$
is the off-diagonal long-range order parameter.
Note that the constant $C$ is absent for the case $\bar n=0$ in which
the correlation functions show pure power-law behavior $G(x,0)\sim x^{-1}$
\cite{Cha00,Alet03}.
Figure~\ref{fig:offd} shows the scaling behavior of $|\langle\psi\rangle|$.
We obtain the mean-field critical exponents $\nu=0.5$ and ${\beta}/\nu=1$,
where ${\beta}$ is the critical exponent characterizing the scaling
properties of the order parameter.

\begin{figure}
\center
\includegraphics[width=2.9in]{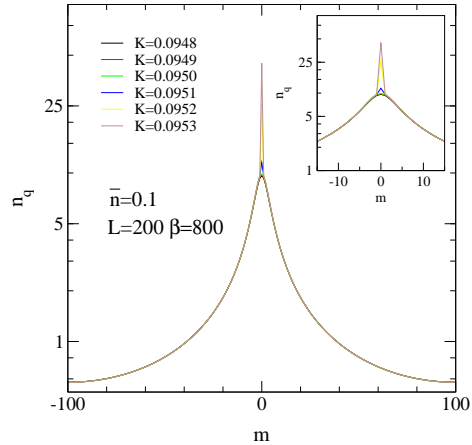}
\caption{Momentum distribution of the quantum rotor model near the superfluid-insulator
transition. The density of condensate at $\vec q=0$ rises sharply as the superfluid
onset transition is tuned.
}
\label{fig:mom}
\end{figure}

The long-range off-diagonal order is coming from the single-particle
condensation. We can find this clearly by observing the momentum
distribution, $n_q$, in Figure~\ref{fig:mom}.
Here we plot the momentum distribution as a function of $\vec q=(2\pi/L)(m,m)$.
The density of the single-particle condensate at $\vec q=0$ state rises
as the superfluid onset transition is tuned.

\begin{figure}
\center
\includegraphics[width=2.9in]{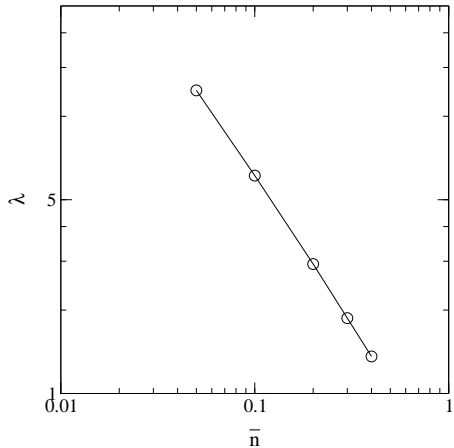}
\caption{The range of the correlation due to the particle-hole excitations
as a function of the charge-offset $\bar n$.
It diverges like $\lambda \sim 0.61/{\bar n}$ as ${\bar n} \to 0$.
}
\label{fig:lambda}
\end{figure}

The momentum distribution consists of the single-particle condensate at $\vec q=0$
and background normal fluid, bringing the macroscopic phase coherence and contributing
to short-range correlations, respectively.
It is quite interesting to note that the normal fluid of interacting bosons is not
the single-particle excitations but the particle-hole pair excitations
which contribute to the asymptotic correlation $G(x,0)\sim x^{-y} e^{-x/\lambda}$
with $y=1$ rather than $y=2$.
The correlation range of the particle-hole pair fluctuations at the transition
diverges like $\lambda \sim 0.61/{\bar n}$ as ${\bar n} \to 0$,
as shown in Figure~\ref{fig:lambda},
ultimately yielding the power-law correlations at $\bar n=0$.

The temporal correlation functions, however, show the power-law behavior
$G(0,\tau)\sim 1/\tau^{(z+\eta)/z}$ with $(z+\eta)/z=1.0$,
implying $\eta\approx 0.0$. 
This exponent and other exponents obtained above confirm
the hyperscaling relation $2\beta=\nu(d+z-2+\eta)$.
But it is not possible to determine
the value of $z$ directly from the correlation functions if they
do not show the power-law behavior.

\begin{figure}
\center
\includegraphics[width=2.9in]{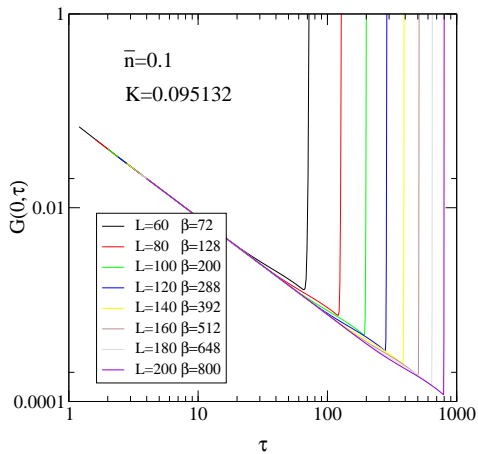}
\caption{The temporal correlation functions at $K=0.095132$ for $\bar n=0.1$.
They show the power-law behavior $G(0,\tau)\sim 1/\tau^{(z+\eta)/z}$ with $(z+\eta)/z=1.0$.
It implies that $\eta\approx 0.0$, but the determination of $z$ is not possible.
}
\label{fig:gt}
\end{figure}

In this work, we have studied
the generic superfluid-insulator transition in interacting Bose systems
via the spherical approximation.
Scaling properties of physical quantities, such as the superfluid stiffness
and the off-diagonal order,
confirm the hyperscaling relations in $d+z$ dimensions,
providing the critical exponent $z=2$, $\nu=1/2$, and $\beta=1/2$.
These values of the critical exponents, supporting the mean-field nature of the
transition, are consistent with theoretical predictions and other calculations.
However, the spatial correlation functions at the transition
do not show the power-law behavior.
Instead they indicate that the correlations consist of two parts:
short-range correlations due to
the normal fluid of the particle-hole pair excitations
and macroscopic long-range off-diagonal order due to the single-particle condensations. 
The temporal correlation function, on the other hand, show the power-law behavior
even though direct determination of $z$ is not possible.

This work was supported by grant No.R05-2004-000-11004-0 from 
Korean Ministry of Science \& Technology.

\end{document}